\documentclass{aa}
\usepackage[dvips]{graphics}
\begin{document}

   \thesaurus{03     
              (11.19.1;  
               11.14.1;  
               11.09.4;  
               09.04.1)}  
   \title{Dust in active nuclei}

   \subtitle{II. Powder or gravel?}

   \author{R. Maiolino,
	   A. Marconi,
          \and
	   E. Oliva
          }

   \offprints{R. Maiolino}

   \institute{Osservatorio Astrofisico di Arcetri
              L.go E. Fermi 5, I-50125, Firenze, Italy\\
              email: maiolino@arcetri.astro.it
             }

   \date{Received ; accepted }

\titlerunning{Dust in active nuclei}

   \maketitle

   \begin{abstract}

In a companion paper, Maiolino et al. (2000) presented various
observational evidences for
``anomalous'' dust properties in the circumnuclear region of AGNs and, in
particular, the reduced $\rm E_{B-V}/N_H$ and $\rm A_V/N_H$ ratios, the
absence of the silicate absorption feature in mid-IR spectra of Sy2s and
the absence of the carbon dip in UV spectra of reddened Sy1s.
In this paper we discuss various explanations for these facts.

The observational constraints favor a scenario where coagulation, catalyzed by
the high densities in the circumnuclear region, yields to
the formation of large grains.
The resulting extinction curve is featureless, flatter than Galactic and the
$\rm E_{B-V}/N_H$ and $\rm A_V/N_H$ ratios are significantly reduced.
These results should warn about an unappropriate
use of the standard Galactic extinction curve and $\rm A_V/N_H$ ratio
when dealing with the extreme gas conditions typical of the circumnuclear
clouds of AGNs.

We also investigated alternative scenarios for the observed
anomalous properties of dust in AGNs. Some of these scenarios
might explain some of the observed properties for a few objects,
but they generally fail to account for all of the observational
constraints obtained for the large sample of AGNs studied in these works.

   \end{abstract}

%

\section{Introduction}

Some authors in the past have found evidences for anomalous properties of the
dust in the circumnuclear region of AGNs. In particular, dust absorption
and reddening appear to be systematically lower than  expected from the
gaseous column density, for a Galactic gas-to-dust ratio and extinction curve
(eg. Maccacaro et al. 1982, Reichert et al. 1985, Granato et al. 1997).
In a companion paper (Maiolino et al. 2000, paper I) we have strengthened these
previous results and provided
additional evidences that the dust in the circumnuclear region of AGNs
 is characterized by anomalous properties.
By comparing the reddening toward the Broad Line Region (BLR) and the gaseous
N$_H$ inferred from the X-rays in a sample of 20 AGNs
we found that, with the exception of three Low Luminosity AGNs,
 the $\rm E_{B-V}/N_H$ ratio is lower than the Galactic standard value by a
factor ranging from $\sim$3 to $\sim$100. By comparing the optical and
X-ray properties of various AGN classes  (type 1.8--1.9 Sys, hard X-ray
selected and radio selected QSOs, Broad Absorption Line QSOs and grism-selected
QSOs) it was inferred that the $\rm A_V/N_H$ ratio must be one or two orders of
magnitude lower than Galactic.
Despite the large mid-IR continuum absorption inferred from the PAHs
 equivalent width, the absence of a significant silicate absorption
feature at 9.7$\mu$m in the average ISO spectrum of Sy2s  indicates
that dust grains in the circumnuclear region
of AGNs have different properties with respect to the Galactic diffuse
interstellar medium.
Finally, the lack of the 2175\AA \ carbon dip in the UV spectra of some type 1
AGNs, which appear affected by dust reddening, suggests a depletion of
small grains in the absorbing medium, as mentioned in paper I.
These findings
suggest that the dust grain distribution is biased in favor of large grains.
The presence of large grains in the circumnuclear region of AGNs was also
suggested by Laor \& Draine (1993) to explain some of the infrared properties
of AGNs.

In this paper
 we examine some possible explanations of the observational findings
presented in paper I. In particular,
we compare our models with the quantitative
constraints given on the $\rm E_{B-V}/N_H$ ratio, the equivalent width of the
silicate feature and the depth of the 2175\AA \ carbon dip.
Note that we often distinguish the $\rm E_{B-V}/N_H$
and $\rm A_V/N_H$ ratios since, when discussing extinction curves different
from Galactic, reddening and absorption are no longer tied by the
standard Galactic
relation $\rm A_V/E_{B-V}=3.1$. In other words, models which could account
for a decreasing of $\rm E_{B-V}/N_H$ do not necessarily lower the
 $\rm A_V/N_H$ ratio, and vice versa.

In Section 2 we first analyze and, in practice, dismiss
scenarios with a standard (Galactic) grain 
size distribution. The case for large grains is analyzed and discussed
in Sect. 3 while in Sect. 4 we draw our conclusions.

\section{The case for standard grains}

\subsection{Metallicity effects}

Various studies have found evidence for super-solar metallicities in the
nuclear region of AGNs (Ferland et al. 1996, Hamann \& Ferland 1993, 1999,
Fosbury et al. 1999, Reynolds et al. 1995, Dietrich \& Wilhelm-Erkens 2000).

The gaseous column densities used in paper I were inferred from the
photoelectric cutoff observed in the X-rays, but actually such
a cutoff gives
a measure of the column of metals N$_Z$ (Morrison \& McCammon
1983), which is then converted into
hydrogen column N$_H$ by assuming solar abundances. If the metallicity is
higher than solar, then the N$_H$ inferred from the X-rays would be
overestimated. On the other hand, a large fraction of refractory elements
is locked into grains. As a
consequence, an increased metallicity would also imply a higher dust content.
Both the dust content and the N$_Z$ to N$_H$ conversion factor depend
linearly on the metallicity and, therefore, any change in metallicity should
not affect the $\rm E_{B-V}/N_H$ ratio in the way we measure it. Finally, we
should emphasize that there are no obvious astrophysical nucleosynthesis
processes that favor the formation of non-refractory elements over refractory
ones, implying that even ``peculiar'' metallicities (in terms of relative
abundances between metals) should not affect the observed
$\rm E_{B-V}/N_H$ ratio.

Also, metallicity effects cannot explain the
discrepancy between the lack of silicate feature and the strong mid-IR
absorption in Sy2s.
In the Small Magellanic Cloud the lack of the 2175\AA \ carbon dip is
ascribed to a low metallicity, but this is unlikely to be the explanation
for AGNs since, as discussed above, generally the latter are characterized
by super-solar metallicities.

Although metallicity effects cannot explain the anomalous properties of dust
in AGNs, it can have important consequences on the expected
emission line fluxes in the case of ionized gas containing dust. This is
an issue which will be discussed further in sections 2.6 and 3.

\begin{table}
\begin{tabular}{lccccc}

Name		& P$_B$		& P$_V$ & P$_H$ & Comm.$^a$ & Refs. \\
\hline
M81		& 		& 0.36	&	&	   & 1 \\
NGC1365		& 1.19		& 0.91	& 0.11	&	   & 4 \\
MCG-5-23-16	& 0.42		& $<$0.8	& 0.7	&	   & 4 \\
NGC5506		& 3.10		& 2.35	& 2.44	& transm.  & 2,5 \\
NGC2992		& 2.62(15.9)	& 3.1	& 1.18	& transm.  & 3,4,6 \\
Mkn6		& 0.5		&	&	&	   & 5 \\
3C445		& 		& 2.1	& 0.75	& P$_{H\alpha}$=1.0 & 4,7 \\
PG2251+113	& 1.0		& 	&	&	   & 8 \\
IRAS1319-16	& 1.1(7.4)	&	&	& P$_{H\alpha}$=0.2 & 2,3 \\
NGC526a		& 0.26		&	&	&	   & 5 \\
Mkn231		& 7.3		& 4.2	& 0.6	& transm.  & 2,9 \\
\hline
\end{tabular}
\caption{Percentage of polarization in the optical (B,V)
 and in the near-IR (H) for the
subsample of the sources listed in Tab.~1 of paper I
for which this information is available.
Polarization fraction in parenthesis are corrected for dilution of the stellar
continuum. Notes: $^a$ {\it transm.}-- objects for which there is
substantial evidence that most of the polarization is due to
 dichroic transmission;
{\it P$_{H\alpha}$}-- gives the polarization percentage
of the broad H$\alpha$ line. References: 1) Barth et al. (1999),
2) Young et al. (1996), 3) Kay (1994), 4) Brindle et al. (1990),
5) Berriman et al. (1989), 6) Goodrich (1992), 7) Corbett et al. (1998),
8) Wills et al. (1992), 9) Goodrich \& Miller (1994).}

\end{table}

\subsection{Reflected broad lines}
Another possibility is that the broad lines used to measure $\rm E_{B-V}$ in
paper I are
not observed through the same medium that absorbs the X-rays (the obscuring
torus), but are scattered by a medium (mirror) observed through a column
density much lower than along the line of sight of the X-ray source.
In this case the nuclei of these objects are expected to show
significant polarization.
Tab.~1 gives the optical and/or near-IR polarization for the objects
examined in paper I (therein Tab.~1) for which this information is available.
For most of them the polarization is $<$2\%, at least in the wavebands
where the reddening is estimated (or as directly
indicated by the polarization of H$\alpha$ in two cases). For NGC5506, NGC2992
and Mkn231, which show higher polarization, there are evidences that most of
the polarization is not due to reflection but to dichroic transmission.

\begin{figure}[h]
\resizebox{\hsize}{!}{\includegraphics{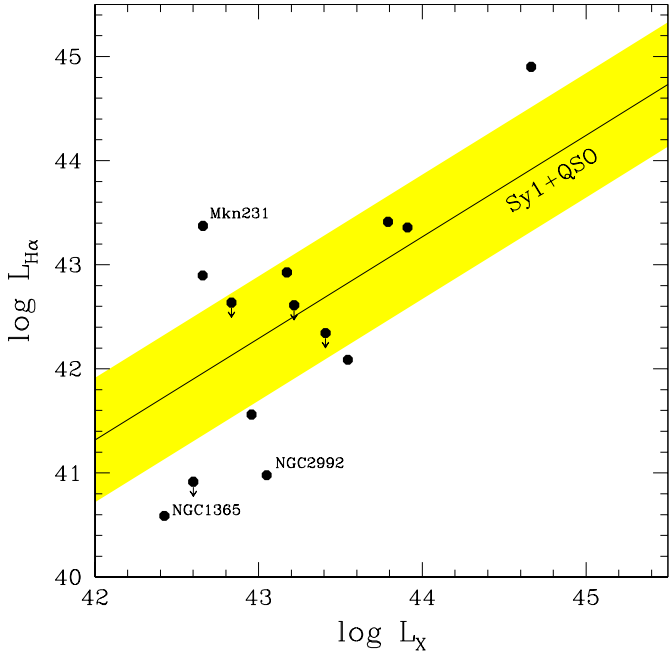}}
\caption{H$\alpha$ luminosity for the objects reported in Tab.~1 of
paper I, corrected
for the reddening derived by assuming a Galactic extinction curve, as
a function of the 2--10 keV luminosity (also corrected for absorption).
The thin line and shaded region show the relation of these two quantities
observed in Sy1 and QSOs (mean and $\pm 1\sigma$ dispersion).
}
\end{figure}

Also, the scattering efficiency generally does not exceed a few percent. As
a consequence, we would expect that the broad lines observed in our objects,
once corrected for the observed reddening, are underluminous by a factor
of $\sim$100 with respect to the intrinsic luminosity indicated by the
absorption-corrected X-ray luminosity, when compared to unobscured type~1
Seyferts and QSOs. Fig.~1 shows the broad
H$\alpha$ luminosity, reddening-corrected
assuming a Galactic standard extinction curve,
versus the intrinsic X-ray luminosity, again for the objects listed in
Tab.~1 of paper I. The solid line and the
shaded region show the L$_{H\alpha}$ vs. L$_X$ relation (mean
and $\pm 1 \sigma$ dispersion respectively) derived for a large
sample of PG QSOs and nearby Sy1s not showing evidence for absorption.
Clearly none of the objects is underluminous in broad $H\alpha$ by the factor
of $\sim 100$ expected in the case of reflection. The most
``$H\alpha$--deficient'' object is NGC2992.
This object does show significant polarization (Tab.~1),
but as discussed above this is very likely a dichroic effect
due to transmission through the host galaxy, which is seen edge-on; an
interpretation which is also supported by the finding that the [OIII] line is
polarized at the same level as the continuum and broad H$\alpha$ line
(Goodrich 1992). As we shall discuss in Sect.~3,
the most likely interpretation for the $H\alpha$ deficiency in this
object is a peculiar extinction curve.
It should also be noted that Mkn231 is actually significantly overluminous in
$H\alpha$, by a factor of $\sim$30. This object is actually known to be
underluminous in the X-rays. Possibly, the absorbing column density might be
much higher than that estimated by Iwasawa (1999) and Turner (1999)
in their faint hard X-ray
spectrum. In the latter case the intrinsic X-ray luminosity could be much
higher.

We should however mention that the sample presented in paper I (Tab.~1)
is probably biased in favor of
objects which are particularly bright in the broad lines emission,
just as a consequence
of the selection criterion (i.e. those absorbed objects which show at
least some broad emission lines). Nonetheless, it is highly unlikely that such
a selection effect accounts for a factor of $\sim$100 in the H$\alpha$ flux.

It is worth noting that a diagram analogous to Fig.~1 was presented
by Veilleux et al. (1999, therein Fig.~3) when discussing the detection
of broad Pa$\alpha$ in Ultraluminous Infrared Galaxies (ULIRGs); there
the broad
H$\alpha$ is replaced by the broad Pa$\alpha$ and the (intrisic) X-ray
luminosity is replaced by the far-IR luminosity. From that diagram 
it can be similarly inferred that the broad component of the Pa$\alpha$
detected in ULIRGs cannot be ascribed to reflection. For what concern Seyfert
galaxies, Veilleux et al. (1997) reach a similar conclusion stating that
the broad component of the hydrogen near-IR lines detected in some Sy2s cannot
be ascribed to reflection.

Finally, also in the reflection scenario
the lack of silicate absorption in Sy2s
and of the lack of the carbon dip in reddened Sy1s is not explained.

\subsection{Reduction of the extinction scattering component}

Unlike the case of Galactic stars where both absorption and scattering
contribute to the extinction, in obscured AGNs we may have
effects of scattering back into our beam. This would imply a reduction
of the scattering component in the effective extinction curve. However, the net
effect must be small compared to what observed for the objects of our sample.
Even in the most extreme case that the absorbing/scattering medium
covers the whole solid angle seeing by the nuclear source (that is very
unlikely) and that the scattered radiation enters our beam without undergoing
any dust absorption (that is also very unlikely), then the effect would be to
completely remove the scattering component out of the total exinction curve.
Since at the wavelengths of interest $\tau _{scat} \approx 2~\tau _{abs} \approx
2/3~\tau _{ext}$, the effect of this extreme case would be to reduce the
extinction and reddening by a factor of 3 at most. While in our sample we
observe several cases for which the reddening deficit is much higher.

The analysis above makes the implicit assumption that the line of
sight is ``typical'', but it is possible that the absorbing column along the
line of sight is lower than the scattering column along the other directions.
In the latter case, and under the extreme assumption discussed above,
 the effective extinction could be reduced further.
Yet, this effect of scattering back into the beam would not explain the lack of
absorption features in the UV and in the mid-IR.

\subsection{Extended BLR}

Several reverberation mapping studies have shown that the BLR has a size
of $\sim$0.01 pc. More specifically $\rm R \approx 0.02~L_{45}^{1/2}$ pc
(Clavel et al. 1991, Peterson 1993, Salamanca et al. 1994). On the other hand,
various pieces of evidence indicate that the obscuring torus has a size
of $\sim$1 pc or larger
(Gallimore et al. 1997, Greenhill \&
Gwinn 1997, Ford et al. 1997, Granato et al. 1997, Maiolino 2000, and
references therein). Since the scale of the latter is so much larger
than both the X-ray source and the BLR, the absorbing medium along
the line of sight should be the same for both. However, for many of the
objects in the sample presented in paper I (Tab.~1)
the broad lines have a width of about 2000 km/s, which
is somewhat narrower than usually observed in type 1 Seyferts and QSOs.
This might imply that in these objects the BLR, or the fraction of it that
we observe, is more extended than assumed and with dimensions comparable
to the obscuring torus. As a consequence, the BLR might
be observed, on average, through a lower absorbing column with respect
to the nuclear
X-ray source.

However, since the broad lines in many of these objects are weak,
very broad wings might have been undetected because either lost in the
noise or confused with the underlying continuum features, thus yielding to
an underestimate of the real width of these lines. Also, in some of these
objects the broad lines were observed to vary on time scales shorter than
a few months,
suggesting a BLR scale smaller than 0.1 pc (Eracleous \& Halpern 1993,
Winkler et al. 1992, Rosenblatt et al. 1994,
Giannuzzo \& Stirpe 1996, Peterson et al. 1984, 1982,
Miller et al. 1985, 1979).

Summarizing, we cannot exclude that for some of the objects in our sample the
BLR is extended enough that the broad lines and the (much smaller) 
hard X-ray source are obscured by columns of gas that are significantly
different. However, this is unlikely to be the case for most of the objects.

Finally, the lack of the silicate feature in the mid-IR spectra of Sy2s and
the lack of the carbon dip in the UV spectra of reddened Sy1s remain
unexplained by the differential absorption scenario discussed in this section.

\subsection{Absorption by BLR clouds}

If one or more BLR clouds are located along the line of sight
toward the X-ray nuclear source then they would contribute to the observed
gaseous absorbing column density (BLR clouds are expected to have 
$\rm N_H \approx 10^{23}
cm^{-2}$), but obviously this absorbing column would not affect the observed
broad line flux. However, should the low $\rm E_{B-V}/N_H$ be a feature common
to most AGNs and should the cause of this effect be absorption by BLR
clouds, then this would require a large covering factor of the BLR clouds,
at least as high as the covering factor of the obscuring medium
\footnote{A covering factor of 80\% for the
obscuring torus is inferred both from the Sy2/Sy1 ratio and from the opening
angle of the light cones observed in Sy2s (Maiolino \& Rieke 1995).},
i.e. $\sim$80\% . On the other hand
the covering factor of the BLR clouds is estimated to be low ($\sim$10\%)
based on the absence of any Ly-edge cutoff in the UV spectrum of QSOs.
Therefore, absorption due BLR clouds might explain some
of the cases of low $\rm E_{B-V}/N_H$, but probably not the majority of them.

Another reason to discard this explanation is related to the lack of
strong and short term
variability of the absorbing N$_H$ in the X-rays. If most of the photoelectric
cutoff observed in the hard X-rays
 is due to a single or a few BLR clouds along the line of sight,
then a strong variation of the $N_H$ would be expected on a time scale
of a few weeks or less,
given the expected size of the BLR clouds ($<0.01$ pc) and their velocity.
However, about 1/3 of the objects in our sample were observed two or more
times in the hard X-rays (eg. Gilli et al. 2000, Malizia et al. 1997)
and the $N_H$ inferred from 
the cutoff does not show the strong variations expected from the motion of a
 BLR cloud
passing on the line of sight. There are a few objects (most
of them not in our sample)
whose hard X-ray monitoring has indeed shown variations
of $\rm N_H$, but such variations are on time scales of years, i.e. consistent
with a distance of the absorber from the nucleus of a few pc (Maiolino 2000).
The only case known to show rapid N$_H$ variability is NGC4151.

Again, the issues related to the silicate feature and to the carbon dip remain
unexplained also in this case.

\subsection{Dust sublimation}

The inner part of the obscuring torus is directly exposed to the UV radiation
emitted by the nuclear source. If the UV flux at this location is large enough,
the dust equilibrium temperature can reach the sublimation value (Laor
\& Draine 1993, Phinney 1989). If dust is sublimated
for a significant fraction of the absorbing gaseous column
this would explain the reduced $\rm E_{B-V}$ with respect
to that expected from the observed N$_H$.
However, this model has problems when compared with the low covering factor
inferred for the ionized gas in the circumnuclear region of AGNs, as discussed
in the following.

The capability of the dust in competing with
the gas in the absorption of UV ionizing photons increases significantly with
the ionization parameter U. Indeed, a highly ionized gas is much more
transparent than dust to the UV ionizing
photons. In particular, for a Galactic gas-to-dust
ratio and dust composition, dust grains are expected to dominate the absorption
of photons beyond the Lyman edge when U is higher than about 10$^{-2}$. At
the inner face of the torus the ionization parameter is expected to be about
0.1 or larger (Netzer \& Laor 1993, Netzer 2000)
and therefore most of the UV ionizing photons
 should be absorbed by the dust which
reprocesses this radiation into near and mid-IR emission, which is actually
observed in the nuclear region of AGNs (eg. Maiolino et al. 1995, Clavel et al.
2000, Maiolino et al. 1998, Thatte et al. 1997, Oliva et al. 1999).
As a consequence, although
the obscuring torus probably subtends as much as 80\% of the solid angle
seen by the nuclear UV source, very little line emission is produced.
The emission lines that would be produced by the torus in absence of
dust should have a width significantly smaller than the broad lines
and close to the width of the narrow lines.
The effect of strong absorption of UV ionizing photons
from dust at the inner face of the
torus has been proposed to explain the marked separation between BLR and NLR
and to explain the low covering factor of the NLR. A more detailed discussion
of this issue is given in Netzer \& Laor (1993), in Laor \& Draine
(1993) and Pier \& Voit (1995).
Here we do not discuss further the implications of dust
absorption for the photoionization models of the circumnuclear gas, but
we use these considerations as a constraint for the properties of the
dust at the inner face of the torus.

If most of the gas in the torus is dust-free, especially at the inner face,
because of dust sublimation, then photons at wavelengths
shortward of the Ly edge should ionize the gas in the
torus and should produce powerful hydrogen lines nearly as narrow as the
narrow lines.
The covering factor of the torus is estimated to be $\sim$80\%.
Until a few years ago the covering factor of the NLR was estimated to
be $\sim$2\% (eg. Netzer \& Laor 1993); however recent observations of the
UV spectrum of QSOs indicate a shortage of ionizing photons with respect to
former estimates by a factor of about 4 (Zheng et al. 1997, Laor et al. 1997)
therefore
 implying a larger covering factor of the NLR ($\sim$8\%), but which
is still much
lower than the covering factor estimated for the obscuring torus.
As a consequence,
the flux of the emission lines produced by the dust-sublimated torus
should be about 10 times larger than the observed narrow lines.

A possible way to suppress the line emission in the dust sublimated medium
is to make the ionized gas very hot. However, in the latter case the
absorbing medium observed in the hard X-rays should appear ``warm'' (according
to the X-ray standards). On the other hand,
in most of the objects of the sample  presented in paper I the absorbing gas
is ``cold'' beyond doubt.

Summarizing, it is difficult to ascribe the low $\rm E_{B-V}/N_H$ ratio to dust
sublimation at the inner face of the torus without running into a serious
emission line budget problem.

For what concerns the absorption features, dust sublimation might explain the
lack of the carbon dip at 2175\AA \ in reddened Sy1s, since the very small
grains which are responsible for this feature undergo sublimation more easily
than larger grains.
However, it is difficult to ascribe the lack of the silicate feature in the
spectra of Sy2s to sublimation,
indeed the absence of this absorption feature
requires a dust distribution dominated by grains larger than
a few $\mu$m (i.e. significantly larger than the maximum dimension of 0.25$\mu$m
inferred for the Galactic medium), rather than
the destruction of small grains (Sect.3.3).

\subsection{Low dust-to-gas ratio}

One of the simplest ways to explain the observed low $\rm E_{B-V}/N_H$
is to assume that the dust-to-gas mass ratio in the obscuring medium
is lower than the standard Galactic value. However, a reduced abundance of dust
in the dense medium of the circumnuclear clouds is not obvious to
justify. The vicinity of the
AGN could help to create the conditions that 
act to destroy dust grains. In particular, sputtering by X--rays
and/or shocks might be a possible mechanisms of dust destruction in this
environment.

\begin{figure}[!h]
\resizebox{\hsize}{!}{\includegraphics{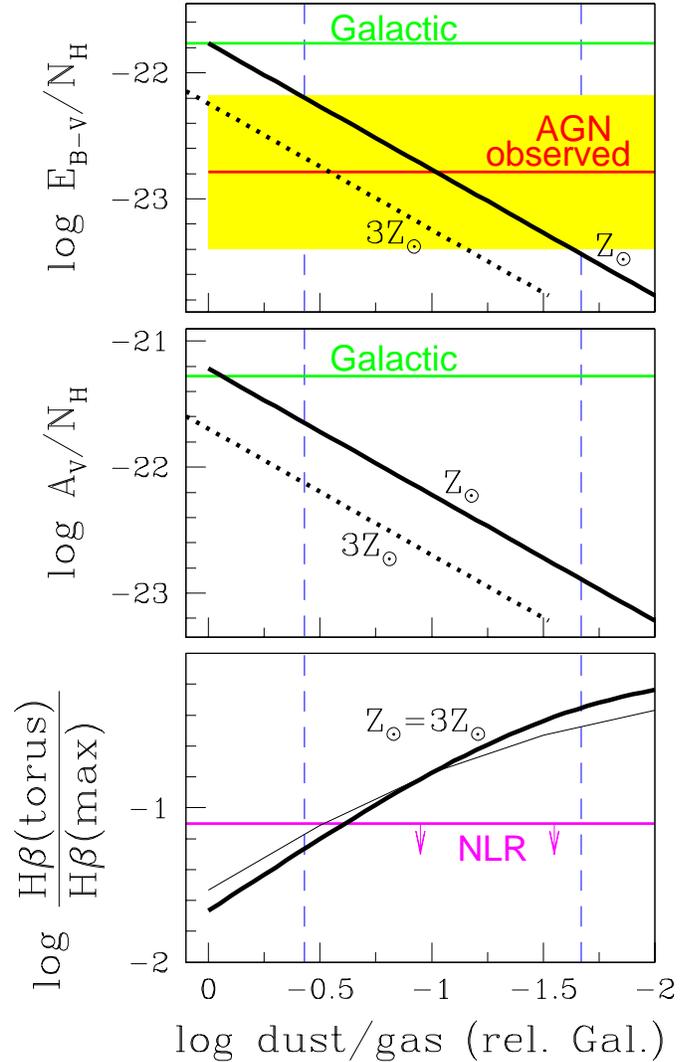}}
\caption{
Expected variation of extinction and emission line properties
as a function of the dust-to-gas ratio relative to the Galactic value.
The predictions of the models, which assume a standard dust size 
distribution, are shown by the thick solid lines.
The upper two panels show the expected variation of the
 $\rm E_{B-V}/N_H$ and $\rm A_V/N_H$ ratios.
The line and shaded area marked by ``AGN observed'' are the mean and
$\pm 1 \sigma$ dispersion value found in the sample of paper I.
The thin dashed vertical lines indicate the range of gas-to-dust ratios that
matches the observed $\rm E_{B-V}/N_H$ for $\sim$68\% of the objects ($\pm 1
\sigma$). The thick dotted lines indicate the effect of an increased
dust-to-gas ratio due to a metallicity three times solar.
In the bottom panel ``H$\beta$(torus)'' is the line luminosity produced
within the dusty torus while ``H$\beta$(max)'' is what one expects if
all the ionizing photons are absorbed by the gas (i.e. the dust-free case).
The light horizontal line constrains the maximum
line emission produced by the torus (in terms of H$\beta$(torus)/H$\beta$(max))
based on the maximum covering factor of the Narrow Line Region ($\sim$8\% ).
The thin solid line in the bottom panel shows, for comparison,
the result obtained with Cloudy.
}
\end{figure}

Obviously, the $\rm E_{B-V}/N_H$ ratio decreases linearly with the
dust-to-gas ratio, if the dust properties remain unchanged. This is shown
in the top panel of Fig.~2. The trend of $\rm E_{B-V}/N_H$ as a function
of the dust-to-gas ratio relative to the Galactic value is given by the
thick solid line. The horizontal thin line and shaded region show the mean
and $\pm 1 \sigma$ dispersion of $\rm E_{B-V}/N_H$ for the objects in the
sample presented in paper I\footnote{In this analysis we excluded the
three Low Luminosity AGNs and the three objects whose reddening
was estimated by means of the continuum fit (paper I).}.
A factor of about 10 lower dust-to-gas ratio is required to match the mean
of the observed $\rm E_{B-V}/N_H$ values.
The thin dashed vertical lines indicate the range of gas-to-dust ratios that
matches the observed $\rm E_{B-V}/N_H$ for $\sim$68\% of the objects ($\pm 1
\sigma$), which implies a dust-to-gas ratio between $\sim 3$ times and
50 times lower than Galactic.
A dust underabundance of factor of $\sim$100 is required to explain the
lowest $\rm E_{B-V}/N_H$ found in paper I (see Fig.~1 in that paper).
The central panel shows
the trend for $\rm A_V/N_H$, which also decreases linearly since the
absorption curve was assumed constant.

As the dust content decreases, the absorption of UV ionizing photons
 by dust grains
 becomes gradually
less important and, as discussed in the previous section, the emission
of nebular
lines gradually increases. In order to check whether the required dust
depletion runs into the same emission line budget problem as in the case of
the sublimated dust (Sect.2.5), we have created a simple semi\---analytical code
to calculate the ionization structure and
hydrogen lines emissivity of a gaseous cloud containing a variable
amount of dust. We assumed a ionization parameter U=0.1
at the inner face of the cloud, as this is about the value expected at
the inner face of the torus (Netzer \& Laor 1993). The covering factor was
assumed to be 80\%.

The results of the model were then compared with the expected emission line
flux in the case of a dust-free medium with a covering factor equal to unity,
i.e. the maximum hydrogen line flux (H$\beta$(max))
which can be obtained with the available photoionizing luminosity.
The ratio between the hydrogen line flux in the dusty 
and in the dust-free cases gives the importance of the line emission from the
obscuring torus relative to other gaseous clouds which are either
free from dust, such as the BLR clouds, or whose ionization parameter is
low enough to make unimportant the effect of dust absorption of
UV ionizing photons, such as
the NLR clouds. The thick black solid line in the bottom panel of
Fig.~2 shows the ratio between
the H$\beta$ flux in the case of a dusty medium (80\% covering factor)
 and the case of a dust-free medium (100\% covering factor),
H$\beta$(torus)/H$\beta$(max),
as a function of the dust-to-gas ratio in the obscuring torus.

We also performed the same calculation with Cloudy, the photoionization code
provided by Ferland and collaborators (Ferland 1999), which accounts for
 several effects which are not included in our simple code. The output of Cloudy
is overplotted with a thin solid line in the bottom panel of Fig.~2. Cloudy's
result is not much different from ours and, partly, it is also to ascribe to
the extinction curve used by Cloudy which is somewhat different from the
Galactic standard curve (eg. $\rm R_V = A_V/E_{B-V} =4$, to be compared
with the standard Galactic value of $\rm R_V = 3.1$). This comparison justifies
us in using our code also for other models in the next section (Sect.3).
The advantage of
our code is that is it possible to use an arbitrary dust composition with a
(quasi-) continuous distribution of grain sizes.
 
The H$\beta$(torus)/H$\beta$(max) values obtained with the model discussed above
have to be compared with the flux of the narrow lines.
More specifically, the flux of the lines emitted by the torus should be lower
than the flux of the narrow lines not to incur in the emission line budget
problems discussed in Sect.2.5; this implies that
H$\beta$(torus)/H$\beta$(max) must be lower than the covering factor of the
NLR clouds, i.e. $<$0.08 (Sect.2.5).
This limit is shown with an horizontal line in the
bottom panel of Fig.~2. The dust-to-gas ratio required to match the
mean observed $\rm E_{B-V}/N_H$ would give
H$\beta$(torus)/H$\beta$(max)$\approx$0.15 that is inconsistent with the
NLR covering factor. Although,
higher values of $\rm E_{B-V}/N_H$ might be marginally
consistent with the NLR emission, the values 
of $\rm E_{B-V}/N_H$ lower than the mean would imply an even
higher H$\beta$(torus)/H$\beta$(max), as shown in Fig.2.

In Sect.~2.1 we mentioned that the nuclear region of AGNs is generally
characterized by super-solar metallicities.
As we are going to discuss, a high metallicity helps
to relax the constraints on the emission
line flux. Metallicity variations change the dust-to-gas
ratio, but the variation of the dust-to-gas ratio due to metallicity leaves
unaffected the {\it measured} $\rm E_{B-V}/N_H$ and
$\rm A_V/N_H$. As discussed in Sect.2.1,
this is because the $\rm N_H$ is measured in the hard X-rays through the
column of metals. On the other hand, the increased dust-to-gas ratio,
regardless of whether due to metallicity or to other effects, does affect
the dust-to-gas {\it opacity} for the UV ionizing photons
 and, therefore, changes the
H$\beta$(torus)/H$\beta$(max) accordingly. As a consequence, even if
metallicity does not affect the $\rm E_{B-V}/N_H$ ratio, it can ``decouple''
the latter ratio from the emission line budget constraints. In particular,
the dotted lines in Fig.~2
show the effect of a metallicity three times higher than solar. On the
top and central panel of Fig.~2 the (metallicity--)
increased dust-to-gas ratio leaves unaffected
the quantities on the Y-axis, thus yielding a shift on the X-axis. Instead,
in the bottom panel
the H$\beta$(torus)/H$\beta$(max) ratio does change accordingly to the
reduced dust-to-gas ratio, i.e. this ratio keep following the trend given
by the solid line. As a result, for the same $\rm E_{B-V}/N_H$ ratio
the implied H$\beta$(torus)/H$\beta$(max) is significantly lower.
In particular,
at the dust-to-gas
ratio which gives an $\rm E_{B-V}/N_H$ accounting for the mean observed value,
H$\beta$(torus)/H$\beta$(max) $\approx$
 0.07, which is marginally consistent with the covering factor of the NLR.
Yet,
for lower values of $\rm E_{B-V}/N_H$ the line emission from the torus is still
much larger than the NLR.

Summarizing, a low gas-to-dust (mass) density ratio might probably explain
the reduced $\rm E_{B-V}/N_H$ for several of the objects without
incurring in emission line budget problems, especially if the
metallicity is super-solar as claimed by several authors. However, low
values of $\rm E_{B-V}/N_H$ (lower than the mean)
are still difficult to explain with this model.

Certainly, the reduced dust--to--gas ratio alone cannot account for the lack of
the silicate feature in the mid-IR spectra of Sy2, nor for the lack of the
carbon dip in reddened Sy1s.

\section{The case for large dust grains}

\subsection{Plausibility of the large grains scenario}

So far we have adopted the standard extinction curve observed in the
diffuse Galactic interstellar medium.
However, the physical conditions of the
gas in the nuclear region of Seyferts and QSOs are so extreme that there is no
reason to assume that dust has the same properties as in the diffuse
interstellar medium of our Galaxy.
The gas densities of the obscuring medium in AGNs probably exceeds
10$^5$--10$^6$ cm$^{-3}$,\footnote{Such a high density is inferred by the
requirement of confining columns of gas higher than $\rm 10^{24}-10^{25}
cm^{-2}$
in a region smaller than a few pc, to avoid a gas mass higher than the
dynamical mass (Risaliti et al. 1999).
The presence of clouds with densities of 10$^5-10^6$ cm$^{-2}$ in the
circumnuclear region of AGNs is also directly
inferred from HCN and other millimetric
observations of AGNs with arcsecond resolution (eg. Tacconi et al. 1994).}
i.e. a density similar or higher than observed in the
densest dark clouds of our Galaxy.
Along the line of sight of dense Galactic molecular clouds
the extinction curve is generally found to be flatter than in
the diffuse ISM (Vrba \& Rydgren 1984, Carrasco et al. 1973, Whittet et al.
1976, Cardelli et al. 1989).
Various observational evidences indicate that most likely such
flat extinction curves are due to a distribution of grain sizes on average
larger than in the diffuse ISM (Kim et al. 1994, Weingartner \& Draine
2000). Larger grains may form either by accretion from
the gas phase or by grain coagulation. The former case should increase the
$\rm A_V/N_H$ ratio, while the latter case should generally yield a reduced
$\rm A_V/N_H$. Although measurements of $\rm A_V/N_H$ are very uncertain, some
works suggest that $\rm A_V/N_H$ is lower in dense clouds characterized by
flatter extinction curves, therefore supporting the coagulation scenario
(Jura 1980, Kim \& Martin 1996). Also, theoretical considerations
indicate that coagulation is much more effective for the growth of dust grains
than accretion from the gas phase (Draine 2000).

Since the grain coagulation rate increases with density ($\propto n^{1/2}$,
Draine 1985), the effect is expected to be even more dramatic in the
circumnuclear clouds of AGNs. In particular, for a gas density in excess
of 10$^6$ cm$^{-2}$ the estimated time scale for the depletion of small
grains due to coagulation is shorter than about 10$^5$ yr, i.e. shorter
than the typical dynamical time scale in the circumnuclear region
of AGNs. This not only implies that dust grains have time
to coagulate before being replaced by gas coming from the outer regions,
but actually some mechanism that prevents grains to grow without limits
must be invoked (eg. shocks).

Summarizing, a dust distribution biased in favor of large grains, probably
as a consequence of coagulation,
is naturally expected in the extreme physical conditions of the gas
in the circumnuclear region of AGNs.

The presence of large grains in the  obscuring medium of AGNs was also proposed
by Laor \& Draine (1993) in relation to the IR properties of active nuclei. In
the following we discuss the effects of large grains, specifically formed by
coagulation, on the observational quantities discussed in paper I.

\subsection{The effect of large grains on $\rm E_{B-V}/N_H$}

To obtain a more quantitative description of the effects of large grains in
the circumnuclear region of AGNs we modelled the
size grain distribution with a power-law:
\begin{equation}
\rm dn(a) \propto a^{-q}~da,~~a_{min}\le a \le a_{max}
\end{equation}
where a is the grain diameter (assumed spherical) and dn(a) the
density of grains with size between a and a+da. Absorption
and scattering cross sections were calculated by using the refractive
indices in Draine \& Lee (1984) and in Laor \& Draine (1994),
and by numerically solving the Mie formulas.
The extinction curve in the Galactic diffuse ISM
was successfully described by Mathis et al. (1977) with
a mixture of graphite (47\%) and silicate (53\%) grains and with
$\rm a_{min} = 0.005\mu m$, $\rm a_{max}=0.25\mu m$ and q=3.5
\footnote{Note that this model does not include the very small grains
(PAHs) which are thought to be responsible for many of the emission feature
in the mid-IR spectrum and probably contribute significantly also to the
absorption dip at 2175\AA \ (Weingartner \& Draine 2000).}.
The effect of large grains was investigated by us in various ways: either by
increasing $\rm a_{max}$, or $\rm a_{min}$, or by decreasing q, or by
a combination of these
possibilities. Coagulation was assumed by imposing the total dust mass to be
constant (relative to the gas mass). The increased average grain size
always gives a flatter extinction curve. As long as $\rm a_{max}$ is larger
than 0.5$\mu$m, coagulation also yields to $\rm E_{B-V}/N_H$ and
$\rm A_V/N_H$ lower than the Galactic standard value. We cannot discuss here
all of the possible cases which give flatter extinction curves. In the
following we discuss
three representative families of models, all of which assume the same
$\rm a_{min} = 0.005\mu m$, but different $\rm a_{max}$ and q.

\begin{figure}[!]
\resizebox{\hsize}{!}{\includegraphics{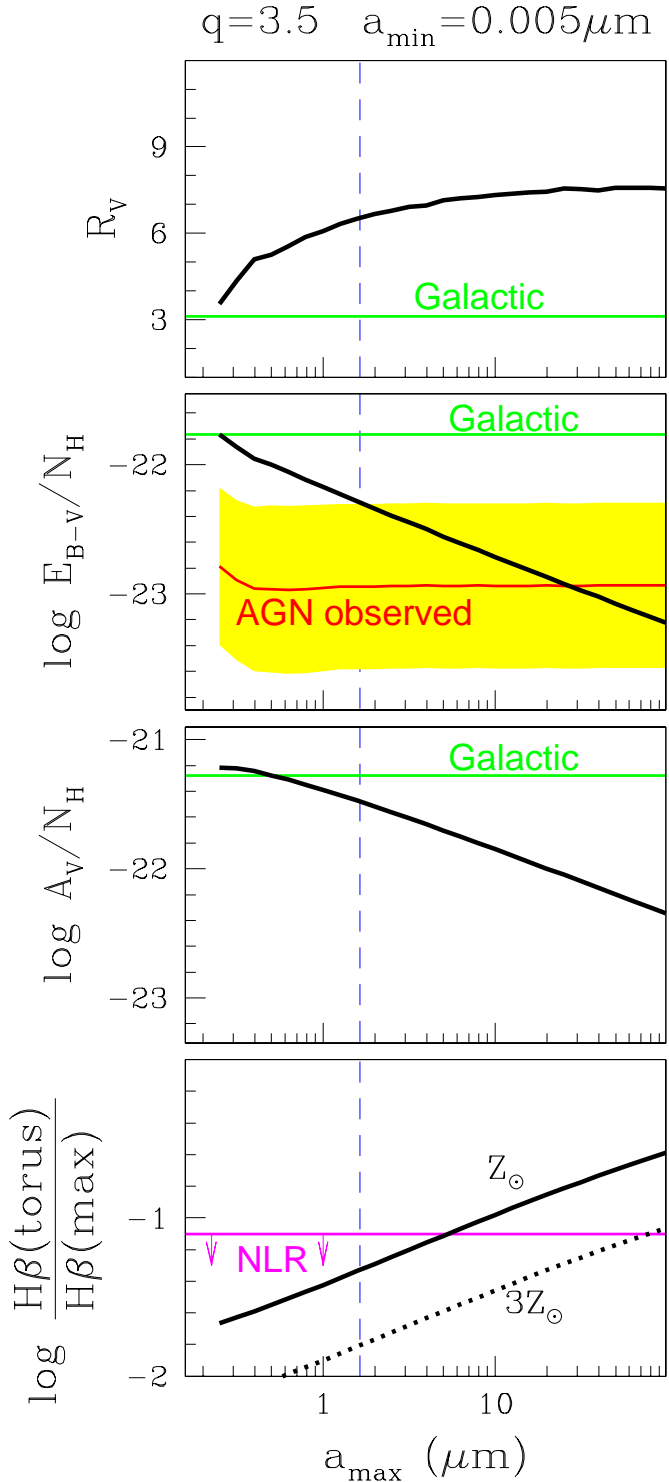}}
\caption{Same as Fig.~2, but in this case the dust-to-gas ratio is constant
while the maximum size of the grain distribution $\rm a_{max}$
varies along
the abscissa. The exponent of the grains size distribution q and the minimum size
$\rm a_{min}$ are kept constant to the standard values of 3.5
and 0.005$\mu$m, respectively. In this figure we also show an additional panel,
the top one, which show the variation of $\rm R_V= A_V /E_{B-V}$.
The thick dotted line in the bottom panel shows the effect of a higher
dust-to-gas ratio due to a metallicity three times solar.
}
\end{figure}

\subsubsection{q=3.5, increasing $\rm a_{max}$}

Fig.~3 is analogous to Fig.~2, but in this case the dust-to-gas (mass) ratio
is kept constant, while $\rm a_{max}$ increases along the abscissa and the
other two dust parameters are kept constant at q$=$3.5 and
$\rm a_{min}=0.005$.

In addition to the panels shown in Fig.~2, we also
include in Fig.~3 a panel giving the variation of $\rm R_V = A_V/E_{B-V}$
(the larger is R$_V$ the flatter is the extinction curve).
In this case R$_V$
increases up to $\sim$7 for very large $a_{max}$. Such large values of R$_V$
imply that the absorption in the optical is higher than that estimated
from the reddening assuming a Galactic extinction curve.

The
H$\alpha$ luminosities for the objects shown in Fig.~1 were corrected assuming
R$_V$=3.1. If R$_V$ is higher than Galactic, then
 the H$\alpha$ luminosities have been
underestimated by the following quantity:

\begin{equation}
\rm \Delta Log(L_{H\alpha}) = {E_{B-V} \over 2.5}~(k R_V-3.1)~0.79
\end{equation}

where $\rm 1 < k < 1.26$ depending on the details of the extinction curve.
For an average $\rm E_{B-V} \approx 0.5$ the correction is significant, i.e.
$\approx 0.7$ in log $L_{H\alpha}$. Some of the objects in Fig.~1
do appear to be that or even more
underluminous in $H\alpha$ with respect to the Sy1+QSO
relation, thus supporting the large grains scenario.
However, other objects are not underluminous in $H\alpha$. For some of them
the required $\rm L(H\alpha )$ correction is small and therefore still
consistent with the Sy1+QSO distribution, but for others the corrected
H$\alpha$ luminosity becomes uncomfortably high ($\sim 2\sigma$ above
the Sy1+QSO distribution). However, as discussed in Sect.2.2 our sample is
probably biased in favor of H$\alpha $ bright sources and, therefore, the
excess of H$\alpha$ luminosity discussed above does not necessary rule out
the large grains hypothesis for these objects.

The second panel of Fig.~3 shows the expected trend for
$\rm E_{B-V}/N_H$ (thick solid line). This ratio has a steep decline as a
function of $\rm a_{max}$. It should be noted that the locus of the values
observed in AGNs (shaded area) is no longer constant as in Fig.~2, this is
because we also took into account that the reddening estimate
$\rm E_{B-V}$ based on the broad line ratios (paper I)
changes with the extinction
curve. The mean $\rm E_{B-V}/N_H$ of the AGNs in our sample can be
explained with $\rm a_{max}\sim 30\mu m$. A much larger value of
$\rm a_{max}$ ($\ge 100\mu m$) is required to match the properties of
objects with lower $\rm E_{B-V}/N_H$.

The ratio $\rm A_V/N_H$ also decreases rapidly with $\rm a_{max}$,
indicating that large grains can, more generally,
explain the low $\rm A_V/N_H$ inferred
for many AGNs and discussed in paper I.

The emission line budget problem does not improve much
with respect to the case
of low dust-to-gas ratio discussed in Sect.2.6: for solar
metallicities, the $\rm H\beta (torus)/H\beta (max)$ ratio ranges from
$\sim$0.04 up to $\ge 0.25$ for the values of $\rm a_{max}$ required to
match the observed $\rm E_{B-V}/N_H$.

In the case of super-solar metallicities the dust-to-gas ratio increases,
but both
the extinction curve, and the $\rm E_{B-V}/N_H$ and $\rm A_V/N_H$ ratios
remain unchanged, as discussed in Sect.2.6. However,  the
$\rm H\beta (torus)/H\beta (max)$ ratio decreases as a consequence of the
increased dust absorption of UV ionizing photons
 (see discussion in Sect.~2.6). The case
of an increased dust-to-gas ratio due to a
metallicity three times higher than solar is
shown with a dotted line in the bottom panel of Fig.~3. In this case the
constraints given by the narrow
 line flux are met better and, in particular, for
a$_{max}$ lower than about 100 $\mu$m,
the expected $\rm H\beta (torus)/H\beta (max)$ is lower than the NLR
constraints. Yet, very low values of $\rm E_{B-V}/N_H$ (observed for
some of the objects analyzed in paper I) remain difficult to
explain with this model.

\subsubsection{$\rm a_{max}=1\mu m$, decreasing q}

In Fig.~4 we show another ``family'' of grain distribution models: here
$\rm a_{max}$ is kept constant and equal to 1$\mu $m, while q decreases
along the abscissa. The extinction curve flattens very rapidly with q,
as also indicated by the rapid growth of $\rm R_V$. As shown, in the
second panel, such flat extinction curve would explain the objects with
nearly no reddening $\rm E_{B-V}$, although substantial gaseous column
$\rm N_H$ is observed along the line (paper I).
However, several of the objects considered in paper I
do show some reddening, therefore indicating that in these objects
the extinction curve cannot be completely flat. A possibility, is that in
the outer parts of the obscuring torus, where the gas density is probably
much lower, the properties of dust are more similar to those of the
diffuse Galactic ISM. In this case the extinction, curve should be
nearly identical to Galactic, but shifted to higher values of absorption
as a consequence of the higher density part of the torus which contributes
with a flat extinction curve. The presence of ``normal'' dust in the outer
parts obviously reduces $\rm R_V$, but also increases $\rm E_{B-V}/N_H$.
The case in which 15\% of the gaseous absorbing column $\rm N_H$ contains
``normal'' dust is shown with a thick
dashed line in Fig.~4. In this case $\rm R_V$
is not unreasonably high, and the $\rm A_V$ inferred from the
observed reddening is still consistent with the presence of weak broad lines
in the optical or IR spectra of the AGNs examined in paper I. Yet, the large
value of $\rm R_V$ makes the $\rm H\alpha$ luminosity
correction even higher than in Sect.3.2.1, which complies with the
H$\alpha$-weak objects, but might be a problem for those objects which are
above the distribution
of Sy1+QSO in Fig.~1 (see analogous discussion in Sect.3.2.1).
Another problem with the inclusion of an even small
(15\%) fraction of dust with Galactic extinction curve is that the ratio
$\rm E_{B-V}/N_H$ has a much slower decline and, as shown in Fig.~4,
cases of very low $\rm E_{B-V}/N_H$ are difficult to explain.

Another important effect of this ``family'' of models is that, at variance with
the $\rm E_{B-V}/N_H$ ratio, the $\rm A_V/N_H$ ratio is not
reduced by a large factor, as shown in Fig.~4. Therefore, this class of models
does not provide the most likely explanation for objects
 characterized by very low $\rm A_V/N_H$ values (paper I).

\begin{figure}[!]
\resizebox{\hsize}{!}{\includegraphics{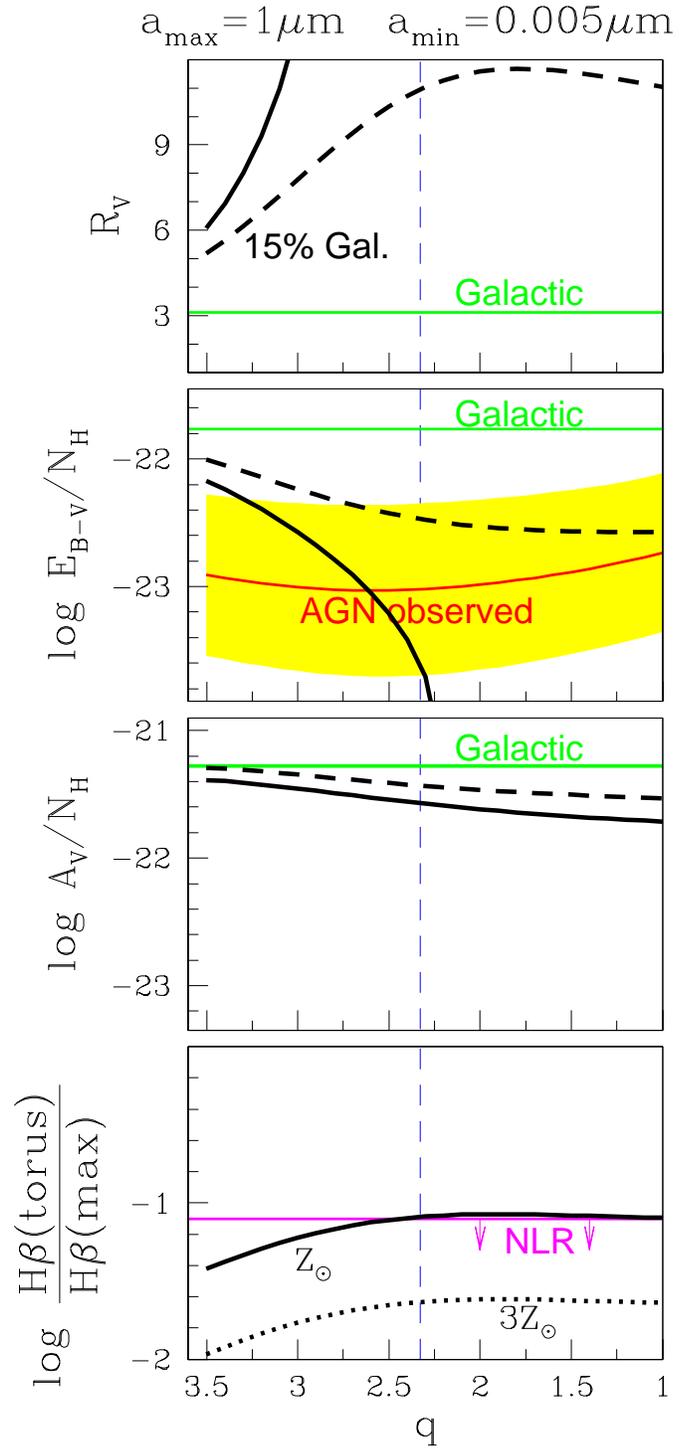}}
\caption{Same as Fig.~3, but in this case the maximum size of the
grain distribution $\rm a_{max}$ is kept constant and equal to 1$\mu$m while
the exponent q is varied along the abscissa. The thick dashed lines show the
case in which 15\% of the gaseous absorbing column (the outer part) contains
Galactic standard dust.
}
\end{figure}

\begin{figure}[!]
\resizebox{\hsize}{!}{\includegraphics{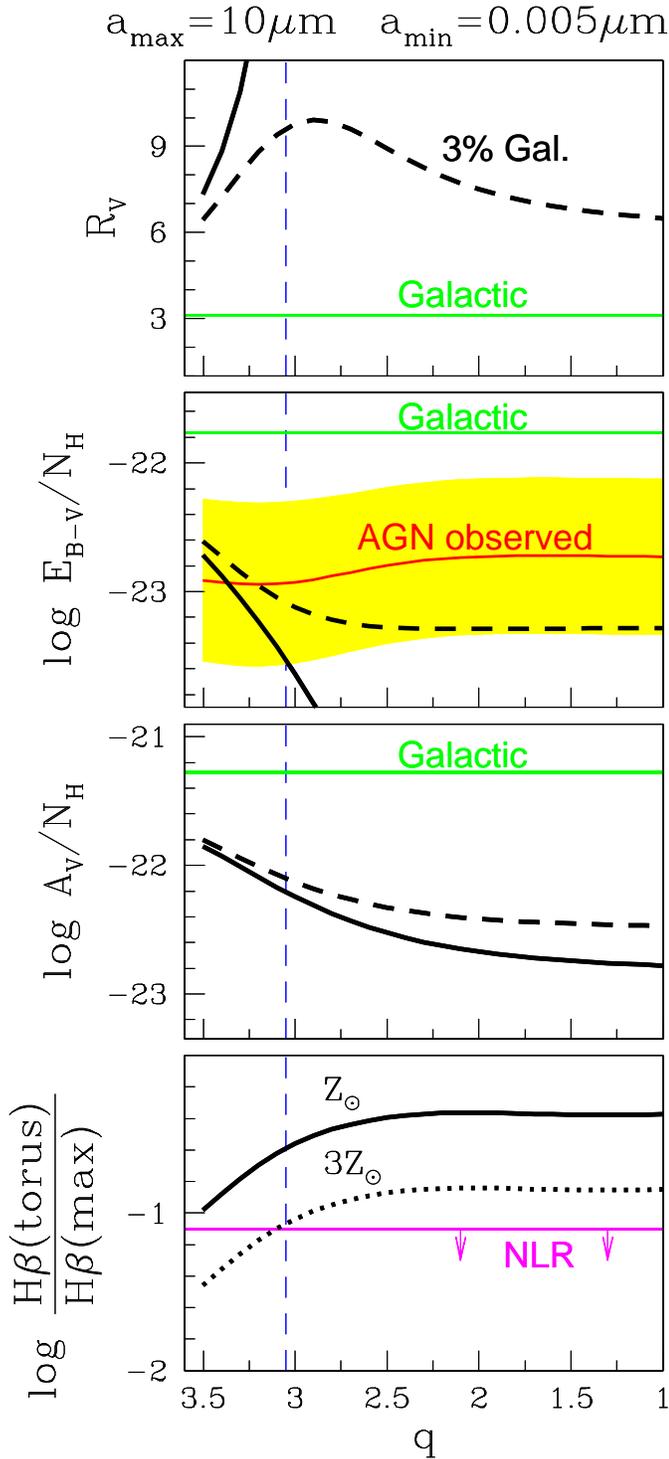}}
\caption{Same as Fig.~4, but with $\rm a_{max} = 10\mu m$.
The thick dashed lines show the
case in which 3\% of the gaseous absorbing column (the outer part) contains
Galactic standard dust.
}
\end{figure}

Yet, one of the most appealing properties of this family of models is that
the $\rm H\beta (torus)/H\beta (max)$ ratio is low ($\sim$8\% at most)
and may be much lower than the NLR constrain, especially
if the dust-to-gas ratio is increased as a consequence of
super-solar metallicity (bottom panel of Fig.~4).
The value of $\rm H\beta (torus)/H\beta (max)$
is obviously unaffected by a fraction of the column density containing
``normal'' Galactic dust, if the latter is located in the outer parts of the
torus.

\subsubsection{$\rm a_{max}=10\mu m$, q decreasing}

This family of models is shown in Fig.~5. These are similar to those
discussed in Laor \& Draine (1993).
Here we do not discuss in detail the properties of these
models, but mostly focus on the main differences and similarities
with the former case.

In these models the effect of decreasing q not only
rapidly flattens the extinction curve, but also decreases significantly the
absorption $\rm A_V$, and therefore can easily explain little reddened objects
with low $\rm A_V/N_H$ (paper I). If the metallicity is three times solar,
or higher, the range of q required to match the $\rm E_{B-V}/N_H$ distribution
of most objects (q$>$3.1)
also complies with the NLR constraint
on the torus emission line flux (bottom panel of
Fig.~5).

As a consequence of the much reduced absorption $\rm A_V$,
the inclusion of an outer layer of
Galactic dust reduces much more $\rm R_V$ and affect much less
$\rm E_{B-V}/N_H$.
This is
 shown by the thick
dashed line in Fig.~5, which gives the case with only 3\% of the
(outer) column of gas
containing Galactic ``standard'' dust. Therefore, this family of
models relaxes the problems on the large $\rm R_V$ discussed in the family
of models discussed in Sect.3.2.2:
it gives very low $\rm E_{B-V}/N_H$ and {\it also}
$\rm A_V/N_H$
values without incurring in large values of $\rm R_V$.

\subsubsection{Summary of the effects on $\rm E_{B-V}/N_H$}

We cannot describe here all of the dust models that matches the specific
constraints for
each single object. However, for most of the AGNs discussed in paper I one or
more dust distribution models match the specific constraints on
$\rm E_{B-V}/N_H$, $\rm E_{B-V}$, broad lines luminosity and detection.

\subsection{The effect of large grains on the silicate feature}

As mentioned in the introduction and discussed more in detail in paper I, Sy2s
appear characterized by significant mid-IR absorption which, nonetheless, is
not associated to a deep silicate absorption feature at 9.7$\mu$m which is
expected by 
a Galactic extinction curve. Since this silicate feature is produced by
grains smaller than a few $\mu$m, a dust distribution
dominated by large grains can also explain this ``anomaly'' of the mid-IR
spectra of Sy2s, as discussed in the following.

\begin{figure}[!]
\resizebox{\hsize}{!}{\includegraphics{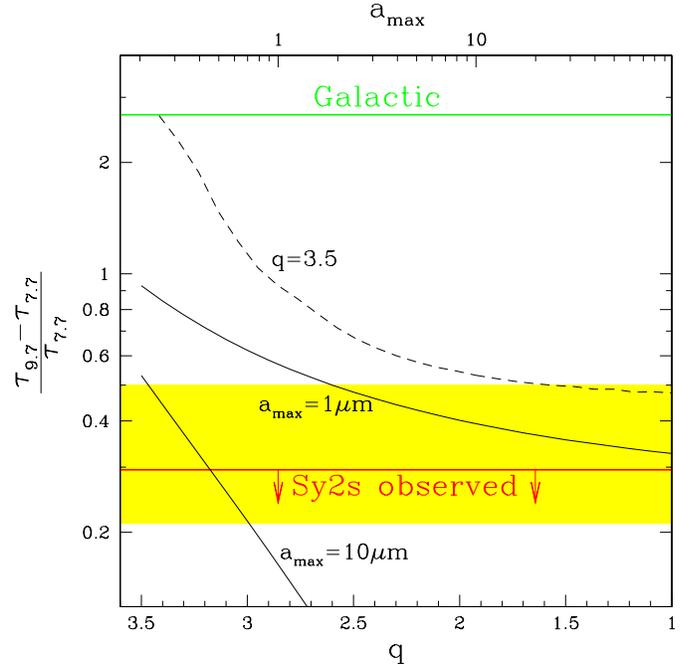}}
\caption{Depth of the silicate feature relative to the featureless mid-IR
absorption as measured by ratio $(\tau_{9.7}- \tau_{7.7})/\tau_{7.7}$. The upper
horizontal line gives the Galactic value, while the lower horizontal line gives
the upper limit obtained for Sy2s (the shaded area gives the spread of the
upper limits). The upper dashed curve gives the ratio expected by the model
presented in Sect.3.2.1
 (q=3.5) and is related to the upper abscissa scale, while
the lower solid curves give the ratio expected by the models presented in
Sects.3.2.2 and 3.2.3
 ($\rm a_{max}=1\mu m$ and $\rm a_{max}=10\mu m$) and are related to
the lower abscissa scale.
}
\end{figure}

We take the optical depth at 7.7$\mu$m as representative of the featureless
absorption in the vicinity of the silicate feature. We specifically choose this
wavelength because is the location of the PAH whose equivalent width is used by
Clavel et al. (2000) to determine the mid-IR absorption. The difference of the
optical depths $\tau_{9.7}- \tau_{7.7}$ is an alternative way to measure the
equivalent width of the silicate feature, as long as it is not saturated.
The ratio $(\tau_{9.7}- \tau_{7.7})/\tau_{7.7}$ is a measure of the depth of the
silicate feature relative to the featureless mid-IR absorption. In Fig.~6 the
upper horizontal line indicates the standard Galactic value of this ratio,
while the lower horizontal line indicates the (mean) upper limit of this ratio
for Sy2s, as inferred from the data in Clavel et al. (2000), and the shaded
area gives the spread of the upper limits (as inferred from the spread of the
EW of the PAH at 7.7$\mu$m). Fig.~6 illustrates in different terms
what already discussed in paper I: the upper limit on the
depth of the silicate feature in the average spectrum of Sy2s is about
one order of magnitude lower than expected from the Galactic extinction curve.
The curves plotted in the diagram show the expected $(\tau_{9.7}- \tau_{7.7})
/\tau_{7.7}$ ratio according to the coagulation
 models discussed in Sect.3.2. In particular, the dashed
line refers to the model presented in Sect.3.2.1 (q constant and equal to 3.5)
and is related to the upper abscissa scale, while the solid curves refer
to the models presented in Sects.3.2.2 and 3.2.3
 ($a_{max}$ constant) and are related to the
lower abscissa scale. All of these models predict a depth of the silicate
feature significantly lower than Galactic. The model with $\rm
a_{max}=10\mu m$ can easily fit the constrain given by the mid-IR
observations, with q just slightly lower than the canonical value of 3.5.

As mentioned in paper I (sect.4.1), the inner (hotter) mid-IR emitting
region probably is not featureless but is also characterized by the 9.7$\mu$m
feature in emission. Since the upper limit on $\tau _{9.7} - \tau _{7.7}$ was
determined assuming an intrinsic featureless continuum, the possible presence
of the 9.7$\mu$m feature in emission should relax such an upper limit.
However, as mentioned in paper I, the equivalent width of the 9.7$\mu$m emission
feature is so small that the upper limit on $(\tau _{9.7} - \tau _{7.7})/\tau
_{7.7}$ would
increase by a quantity of 0.15 at most, which certainly cannot account for the
the discrepancy with the value expected for the Galactic case (Fig.~6).

Finally, we wish to emphasize that the presence of strong PAH features in
the ISO spectra of Seyfert galaxies is not in contractdiction with our large
grain model, since such PAH features most likely come from the host galaxy
(where the dust properties are probably similar to the Galactic ISM) and not
from the nuclear region.

\subsection{The effect of large grains on the carbon dip}

In Sects.3.2 and 3.3
 we only considered the effects of coagulation by increasing $\rm
a_{max}$ and by decreasing q, since these are the parameters which primarily
affect R$_V$, $\rm E_{B-V}/N_H$, $\rm A_V/N_H$ and the silicate feature. The
latter quantities are instead little sensitive to $\rm a_{min}$. However, this
is not the case for the carbon dip which, in the model 
of Mathis et al. (1977) is ascribed to graphite grains with
sizes in the range 100\AA -- 200\AA . In particular, the carbon dip is washed
away from the extinction curve if coagulation depletes small grains by
increasing $\rm a_{min}$ above 200\AA . This is shown in Fig.~7, where we plot
the variation in the UV shape extinction curve by modifying the dust
distribution to have $\rm a_{min} = 0.03\mu m$
and leaving $\rm a_{max}$ and q to the standard values of 0.25 and 3.5,
respectively.

Note that recent studies have shown that most of the profile of the
2175\AA \ dip is probably produced by even smaller grains (PAHs, Weingartner \&
Draine 2000) that are not included in our model.
Obviously, in this case
 the depletion of small grains onto larger grains through coagulation
 would be even more effective in removing the 2175\AA \ dip.

\begin{figure}[!]
\resizebox{\hsize}{!}{\includegraphics{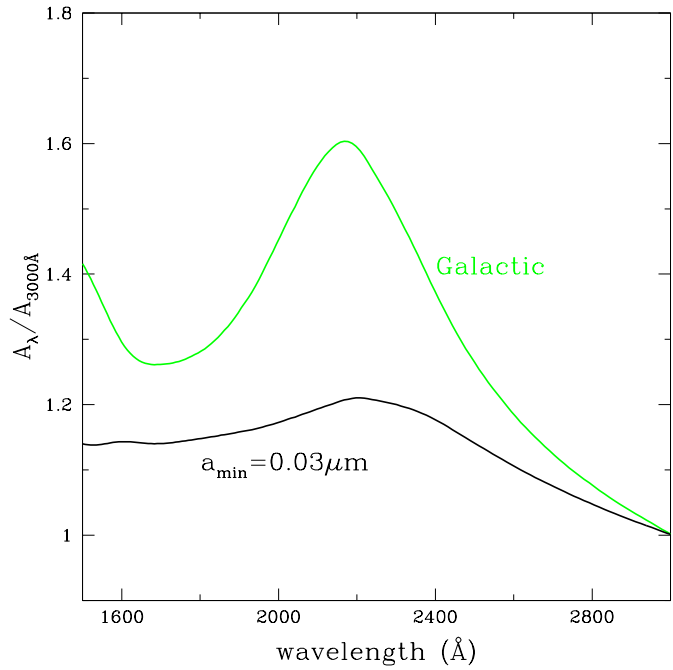}}
\caption{Comparison of the Galactic UV extinction curve with the extinction
curve given by a dusty medium which is depleted of
 grains smaller than 0.03$\mu$m.
Note the absence of the carbon dip in the latter case.
}
\end{figure}

As a consequence, the coagulation scenario is also supported by
the lack of a prominent carbon dip in spectra of reddened Sy1s (paper I).
However, the depletion of small grains can also be achieved with other physical
processes (eg. sublimation, Sect.2.5) and, therefore, the absence of the carbon
dip alone cannot be used to rule out other scenarios. Yet, whatever is the
mechanism responsible for depleting small grains, the absence of the carbon dip
certainly indicates that the properties of dust in AGNs are different from
Galactic and, in particular, that the dust distribution is biased in favor of
large grains.


\section{Conclusions}

We have discussed various interpretations of the observational evidences for
``anomalous'' properties of dust in the circumnuclear region of AGNs and, in
particular, the reduced $\rm E_{B-V}/N_H$, the reduced $\rm A_V/N_H$, the
absence of the silicate absorption feature in the mid-IR spectra of Sy2s and
the absence of the carbon dip in the UV spectra of reddened Sy1s, which are
reported in a companion paper (Maiolino et al. 2000, paper I).

A dust distribution dominated by large grains as a consequence of coagulation
is the interpretation which best fit the various observational constraints.
The formation of large grains is naturally expected
in the high density environment characterizing
the circumnuclear region of AGNs. A dust grain distribution biased in favor of
large grains makes the extinction curve flatter than Galactic and featureless
(especially for what concerns the silicate mid-IR feature and the UV carbon
dip). Coagulation also yields a reduced $\rm E_{B-V}/N_H$ and $\rm A_V/N_H$.

It should be noted that nearly all of the papers, published so far, dealing
with the
effects of dust in AGNs have adopted a standard Galactic extinction curve
(essentially with the only exception of Laor \& Draine 1993, who first
proposed the large grain scenario). The assumption that the properties of dust
in the extreme conditions of the circumnuclear region of AGNs are similar to
the Galactic diffuse interstellar medium is somewhat naive, given that even in
the denser clouds of our Galaxy the extinction curve is already different
with respect to the diffuse medium.
Probably, many of the past studies considering the effects of
dust obscuration and scattering in AGNs should be revised
at the light of the findings obtained in this paper.

With regard to the scattered radiation (especially in obscured AGNs) it is
worth noting that a dust distribution dominated by large grains would yield a
nearly gray reflection, thus mimicking the scattering by free electrons. Large
grains produced by dust coagulation would have a reduced scattering efficiency
with respect to the standard Galactic dust mixture. However, while in some
cases coagulation can result in a reduction of the scattering efficiency by
 more than an order of
magnitude, for some of the dust grain distributions producing a flat
extinction and scattering curve the reduction in the scattering efficincy is
much lower (e.g. for the model discussed in 3.2.2 the scattering efficiency is
reduced by a factor of $\sim$2). Therefore, it is possible that some of the
gray--reflected spectra found in some obscured AGNs, and ascribed to electron
scattering, might actually be produced by large grains scattering.

We have also examined other possible interpretations for the ``anomalous'' dust
properties discussed in paper I, and more specifically:\\

1) a metallicity higher than solar (as often found in the central region
of AGNs) might affect the estimate of the equivalent N$_H$ from the
photoelectric cutoff in the hard X-rays;\\
2) the observed broad lines might be scattered by a reflecting medium through a
path of much lower gaseous column than the X-rays;\\
3) reduction of the scattering component of the extinction curve as a
consequence scattering of the radiation into our beam;\\
4) the BLR might have a size comparable to that of the absorbing medium
(at variance with what assumed by the standard unified model) and, therefore,
suffer less absorption with respect to the nuclear X-ray source;\\
5) the hard X-rays might be partly absorbed by the same clouds which produce the
broad lines;\\
6) the sublimation of dust at the inner face of the obscuring torus would
produce a gaseous dust-free region;\\
7) the absorbing medium might simply be characterized by a low dust--to--gas
ratio (in terms of mass).

Although some of these cases might apply to some of the objects examined in
paper I, none of them can fit all of the observational constraints for most of
the objects, at variance with the large grains scenario.

\begin{acknowledgements}
We are grateful to B. Draine for enlightening discussions during the early
stages of this work.
This paper also benefits of useful comments from
A. Natta, M. Walmsley and from the referee R. Antonucci.
This work was partially supported by the
Italian Space Agency
(ASI) under grant ARS--99--15 and by the Italian Ministry for
University and Research (MURST) under grant Cofin98--02--32.
We thank G. Ferland for making Cloudy available to the community.
\end{acknowledgements}

\end{document}